\documentclass[prb,twocolumn,superscriptaddress,floatfix,showkeys]{revtex4}
\usepackage{graphicx}
\usepackage{amsmath,subfigure,epsfig,psfrag}
\usepackage{comment}
\usepackage{commath}
\usepackage{graphicx}
\usepackage{amssymb}
\usepackage{xcolor}
\usepackage{soul}
\usepackage{listings}
\usepackage{rotating}
\usepackage{latexsym}
\usepackage{amsfonts}
\usepackage{amssymb}
\usepackage{wasysym}
\usepackage{xfrac}
\usepackage{mathtools}
\usepackage{dsfont}
\usepackage{tikz}
\usepackage{placeins}
\usetikzlibrary{matrix,shapes,arrows,positioning,chains}
\makeatletter
\renewcommand*\env@matrix[1][\arraystretch]{%
  \edef\arraystretch{#1}%
  \hskip -\arraycolsep
  \let\@ifnextchar\new@ifnextchar
  \array{*\c@MaxMatrixCols c}}
\makeatother
\newcommand{\SUMP}[2]{\relax\ensuremath{\sum\limits_{#1}^{#2}\hspace{-1.5em}{\phantom{\sum}}^\prime}}
\begin{document}

\title{Efficient Calculation of Derivatives of Integrals in a Basis of Non-Separable Gaussians Through Exploitation of Sparsity}

\author{Jacques K. Desmarais}
\email{jacqueskontak.desmarais@unito.it}
\affiliation{Dipartimento di Chimica, Universit\`{a} di Torino, via Giuria 5, 10125 Torino, Italy}

\author{Alessandro De Frenza}
\affiliation{Dipartimento di Chimica, Universit\`{a} di Torino, via Giuria 5, 10125 Torino, Italy}

\author{Alessandro Erba}
\email{alessandro.erba@unito.it}
\affiliation{Dipartimento di Chimica, Universit\`{a} di Torino, via Giuria 5, 10125 Torino, Italy}

\date{\today}
\begin{abstract}
A computational procedure is developed for the efficient calculation of derivatives of integrals over non-separable Gaussian-type basis functions, used for the evaluation of gradients of the total energy in quantum-mechanical simulations. The approach, based on symbolic computation with computer algebra systems and automated generation of optimized subroutines, takes full advantage of sparsity and is here applied to first energy derivatives with respect to nuclear displacements and lattice parameters of molecules and materials. The implementation in the \textsc{Crystal} code is presented and the considerably improved computational efficiency over the previous implementation is illustrated. To this purpose, three different tasks involving the use of analytical forces are considered: i) geometry optimization; ii) harmonic frequency calculation; iii) elastic tensor calculation. Three test case materials are selected as representatives of different classes: i) a metallic 2D model of the Cu (111) surface; ii) a wide-gap semiconductor ZnO crystal, with a wurtzite-type structure; and iii) a porous metal-organic crystal, namely the ZIF-8 Zinc-imidazolate framework. Finally, it is argued that the present symbolic approach is particularly amenable to generalizations, and its potential application to other derivatives is sketched.

\end{abstract}

\maketitle

\section{Introduction}

Atom-centered Gaussian-type functions (GTFs) were proposed for variational wavefunction calculations in quantum chemistry, independently by Boys \cite{boys1950electronic} and McWeeny,\cite{mcweeny1950gaussian} and nowadays represent an important class of basis functions for practical first-principle calculations. Other notable choices are Slater functions,\cite{slater1930atomic} used in the \textsc{Adf} program, numerical atomic orbitals, used in the \textsc{OpenMX} and \textsc{Siesta} programs,\cite{openmx,garcia2020siesta} or wavelet basis sets used in the \textsc{BigDFT} program.\cite{ratcliff2020flexibilities} For the special case of infinite, periodic, three-dimensional systems, plane waves represent another notable alternative.\cite{VASP2,blaha2020wien2k,giannozzi2017advanced}

In the overwhelming majority of Gaussian-based quantum chemical programs (with \textsc{Crystal} being an exception), integrals are calculated in the basis of so-called Cartesian GTFs (CGTFs), $C_{t,u,v}$, which read:\cite{aidas2014d,DIRAC19,TURBOMOLE,ReSpect-5.1.0,valiev2010nwchem,neese2018software,g16,MOLPRO_brief,aquilante2016molcas}
\begin{equation}
\label{eqn:CGTF}
C_{t,u,v} (\alpha,\mathbf{r}-\mathbf{A})  = (r_x - A_x)^t (r_y - A_y)^u (r_z - A_z)^v e^{- \alpha \vert \mathbf{r} - \mathbf{A} \vert^2} \; ,
\end{equation}
where $t,u,v$ are positive integers, $\mathbf{r}$ is the coordinate of an electron, and $\mathbf{A}$ the center of the basis function (usually the position of an atomic nucleus). A CGTF in Eq. (\ref{eqn:CGTF}) is, then, a \textit{separable} Gaussian as it may be written as a product of three functions:
\begin{equation}
C_{t,u,v} (\alpha,\mathbf{r}-\mathbf{A}) = \prod_i c^{(i)} (\alpha,r_i-A_i) \; ,
\end{equation}
where $i=x,y,z$ is a Cartesian index and
\begin{equation}
c^{(i)}(\alpha,r_i-A_i) = (r_i - A_i)^{T_i} e^{- \alpha (r_i - A_i)^2} \; ,
\end{equation}
with $T_i=t,u,v$ for $i=x,y,z$, respectively. The separability of CGTFs, then, considerably simplifies the computation of integrals. Powerful algorithms based on CGTFs have been developed by McMurchie and Davidson (MD), based on recursion relations of the CGTF pair product.\cite{MurDav} Notwithstanding, the exact order in which to perform the MD recursions is not obvious and any departure from ideality can result in significant loss of computational efficiency.\cite{wilson2003handbook} A variety of ``recursion trees'' have correspondingly been proposed.\cite{johnson1991exact,cisneros1993improved,gill1989efficient} A notable alternative to the MD strategy is the prescription of Obara and Saika, where recursions are developed instead on individual integrals, rather than on CGTF pair products.\cite{obara1986efficient} For the specific case of electron-nuclear attraction and electron-electron repulsion integrals, Dupuis, Rys and King introduced efficient quadrature formulas.\cite{dupuis1976evaluation}

Despite the obvious simplifying advantages of separable Gaussians, CGTFs are not eigenfunctions of the electronic angular-momentum operator, and thus classification based on conventional quantum numbers becomes ambiguous. Therefore, practical quantum-chemical calculations are often instead based on the \textit{non-separable} real solid spherical harmonic GTF (RSSHGTF) functions:
\begin{equation}
\label{eqn:RSSHGTF}
R \left( \alpha,\mathbf{r}-\mathbf{A},n,l,m_l \right)  = \vert \mathbf{r}-\mathbf{A} \vert^{2n} X \left( \mathbf{r}-\mathbf{A},l,m_l \right) e^{-\alpha \vert \mathbf{r}-\mathbf{A}\vert^2} \; ,
\end{equation}
where $n, l, m_l$ are the usual principal, azimuthal and magnetic quantum numbers and $X$ is an unnormalized real spherical harmonic. Although only $n=0$ RSSHGTFs are used as basis functions, the $n \ne 0$ ones are useful as auxiliary functions for computing integrals. If the basis functions are $R$, a calculation of integrals in the CGTF basis requires a subsequent transformation to RSSHGTFs, which may be achieved via:
\begin{equation}
\label{eqn:RSSH}
\vert \mathbf{r} \vert^{2n} \ X \left( \mathbf{r},l,m_l \right) = \SUMP{t,u,v}{} D_{t,u,v} \left( l,m_l \right) r_x^t r_y^u r_z^v \; ,
\end{equation}
where $D_{t,u,v}$ are linear coefficients, and the prime over the sum indicates that it is restricted to triplets $t,u,v$ that satisfy the equality $t+u+v=l+2n$.\cite{Pisani} A more direct and efficient strategy was proposed by Saunders, who suggested to evaluate the integrals directly in the RSSHGTF basis.\cite{Saund} This strategy has been implemented in the \textsc{Crystal} program, alongside powerful screening algorithms and a particularly efficient strategy for evaluating the Coulomb series of infinite-periodic systems, based on Ewald summation and by approximating the Coulomb potential by a distributed point multipole model.\cite{saunders1992electrostatic,Pisani} The approach has also been extended to analytical first energy gradients w.r.t. nuclear displacements and cell parameters.\cite{Doll1,Doll2,DOLL_CELLGRAD1,DOLL_CELLGRAD2} On the other hand, the added complication resulting from the non-separability of RSSHGTFs means, for instance, that second analytical derivatives are not yet available. And the algorithm was only recently generalized to $l=4$ $g$-type functions.\cite{gorbs}

Here we provide a way forward through efficient calculation of derivatives of integrals in a basis of non-separable RSSHGTFs by symbolic computation with computer algebra systems. Our approach is inspired by previous work of Saunders et al. on the calculation of derivatives of the Boys' function.\cite{scott1997numerical} In the case of first energy derivatives, the approach is shown to yield significant improvements over the previous implementation. Generalization to other derivatives of particular interest (second order nuclear derivatives and first-order magnetic field derivatives with field-dependent GTFs) is discussed.

\section{Formal and Computational Aspects}

In the Saunders scheme, the RSSHGTF pair product (or its derivatives) is expanded into so-called Hermite GTFs $\Lambda$:\cite{Saund}
\begin{equation}
\label{eqn:HGTF}
\Lambda_{t,u,v} \left( \alpha,\mathbf{r}-\mathbf{A}\right) = \left( \frac{\partial}{\partial A_x} \right)^t \left( \frac{\partial}{\partial A_y} \right)^u \left( \frac{\partial}{\partial A_z} \right)^v e^{-\alpha \vert \mathbf{r}-\mathbf{A}\vert^2} \; .
\end{equation}
For calculating the integrals themselves, the expansion of the pair product of two RSSHGTFs involves linear coefficients $E$:
\begin{eqnarray}
\label{eqn:RRL}
R \left( \alpha,\mathbf{r}-\mathbf{A},n,l,m_l\right) R \left( \beta,\mathbf{r}-\mathbf{B},n^\prime,l^\prime,m_l^\prime\right) = \nonumber \\
\sum_{t,u,v}^{\mathcal{E} \left( n,n^\prime,l,l^\prime \right) } E_{t,u,v} \left[ n,l,m_l,n^\prime,l^\prime,m_l^\prime \right] \Lambda_{t,u,v} \left( \gamma,\mathbf{r}-\mathbf{P}\right) \; ,
\end{eqnarray}
where the sum over $t,u,v$ runs over all values in the set of integer triplets $\mathcal{E} \left( n,n^\prime,l,l^\prime \right)$ that satisfy  the criteria $t+u+v \le 2n +2n^\prime + l + l^\prime$,  as well as $t \ge 0$, $u \ge 0$, $v \ge 0$. In Eq. (\ref{eqn:RRL}),  $\gamma=\alpha + \beta$ and $\mathbf{P}$ is the centroid of the RSSHGTF pair $\mathbf{P}= \left( \alpha \mathbf{A} + \beta \mathbf{B} \right)/\gamma$.

\subsection{First-Order Derivatives with respect to Atomic Positions}

For the derivative w.r.t. the $i$-th Cartesian component of $\mathbf{A}$, the expansion is done through linear coefficients $G^{A_i}$:
\begin{eqnarray}
\label{eqn:RRLG}
\frac{\partial}{\partial A_i} R \left( \alpha,\mathbf{r}-\mathbf{A},n,l,m_l\right) R \left( \beta,\mathbf{r}-\mathbf{B},n^\prime,l^\prime,m_l^\prime\right) = \nonumber \\
\sum_{t,u,v}^{\mathcal{G} \left( n,n^\prime,l,l^\prime \right) } G^{A_i}_{t,u,v} \left[ n,l,m_l,n^\prime,l^\prime,m_l^\prime\right] \Lambda_{t,u,v} \left( \gamma,\mathbf{r}-\mathbf{P}\right) \; ,
\end{eqnarray}
where the set $\mathcal{G} \left( n,n^\prime,l,l^\prime \right)$ includes all positive integer triplets $t,u,v$ that satisfy  $t+u+v \le 2n +2n^\prime + l + l^\prime+1$. 

The two sets of coefficients introduced in Eqs. (\ref{eqn:RRL}) and (\ref{eqn:RRLG}) are related by:\cite{Doll2}
\begin{eqnarray}
\label{eqn:EG}
 G^{A_i}_{t,u,v} \left[ n,l,m_l,n^\prime,l^\prime,m_l^\prime\right] = \frac{\partial}{\partial A_i} E_{t,u,v} \left[ n,l,m_l,n^\prime,l^\prime,m_l^\prime\right] \nonumber \\
+ \frac{\alpha}{\gamma} E_{t-\delta_{i,x},u-\delta_{i,y},v-\delta_{i,z}} \left[ n,l,m_l,n^\prime,l^\prime,m_l^\prime \right] \; ,
\end{eqnarray}
where $\delta_{i,j}$ is the Kronecker delta. The full set of $E$ and $G^{A_i}$ coefficients may be obtained from a set of recurrence relations, deriving from the corresponding recurrences for spherical harmonics and Hermite polynomials.\cite{Saund,gorbs,Doll2} Once they are known, the coefficients required for derivatives w.r.t. all other centers may be determined as:\cite{Doll2}
\begin{eqnarray}
\label{eqn:centerB}
G^{B_i}_{t,u,v} \left[ n,l,m,n^\prime,l^\prime,m^\prime \right] = - G^{A_i}_{t,u,v} \left[ n,l,m,n^\prime,l^\prime,m^\prime\right]\nonumber \\
+E_{t-\delta_{i,x},u-\delta_{i,y},v-\delta_{i,z}} \left[ n,l,m,n^\prime,l^\prime,m^\prime\right] \quad \forall i=x,y,z \; .
\end{eqnarray}

\begin{table}[b!]
\begin{center}
\caption{Ratio of vanishing/total coefficients required for computing the integrals (E) and their derivatives (G) for RSSHGTF pair product shells of increasing angular quantum numbers. Percentages of vanishing coefficients are also reported in parentheses.}
\label{tab:zeros}
\vspace{5pt}
\begin{tabular}{lcccccccc}
\hline
\hline
$l$-$l^\prime$ or $l^\prime$-$l$ & $s$-$s$ &  $p$-$p$ &  $d$-$d$ & $f$-$f$ \\
&&\\
$G$ & $\frac{6}{12}$ (50\%) & $\frac{372}{540}$ (69\%)  &  $\frac{2470}{4200}$  (59\%) &  $\frac{8926}{17640}$ (51\%)  \\
&&\\
$E$ & $\frac{0}{1}$ (0\%) & $\frac{57}{90}$ (63\%) &  $\frac{486}{875}$ (56\%) &  $\frac{1987}{4116}$ (48\%)
&&\\
&&\\
\hline
\hline
\end{tabular}
\end{center}
\end{table}

\noindent For quantum chemical computations, the application of recurrence relations (i.e. the direct approach) to compute the $E$ and $G^{A_i}$ turns out to be impractical, especially because the ``best'' order in which the recurrences need to be applied for fast computations is not known. Indeed, even in the simpler case of separable Gaussians, combinatorial complexity is substantial, and the optimal algorithm is only known for low quantum numbers.\cite{johnson1991exact} In the case of non-separable Gaussians, the direct application of recurrence relations requires evaluation of a very large number of logical statements, whose cost can be prohibitive.\cite{gorbs} Finally, the direct approach is not well suited for exploiting the sparsity of the $G^{A_i}$.

Indeed, a large number of $E$ and $G^{A_i}$ coefficients vanish from the requirement that integer triplets $t,u,v$ belong to the sets $\mathcal{E} \left( n,n^\prime,l,l^\prime \right)  $ or $\mathcal{G} \left( n,n^\prime,l,l^\prime \right)  $. The importance of sparsity in the computation of $G^{A_i}$ and $E$ coefficients in the $n=n^\prime=0$ case is discussed with the help of Table \ref{tab:zeros}. The table provides the ratio of vanishing/total $G^{A_i}$ and $E$ coefficients for RSSHGTF pair product shells of increasing quantum numbers. In the case of $G^{A_i}$, more than half of the coefficients are vanishing up to $l=3$ $f$-$f$ products. Proper exploitation of sparsity, then, becomes key for efficient computations.

\begin{center}
\begin{figure}[t!]
\centering
\includegraphics[width=8.6cm]{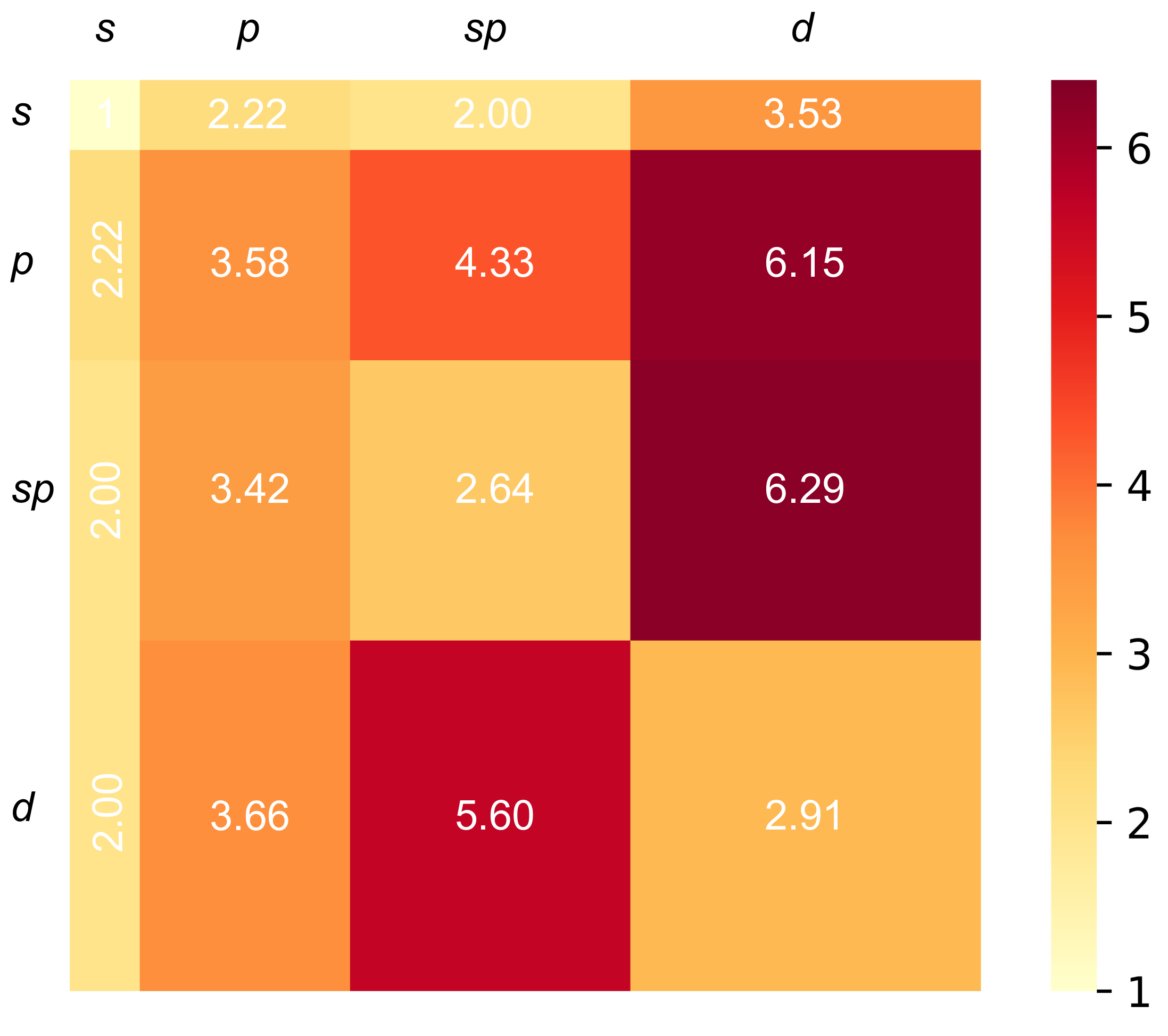}
\caption{Speedups for calculating $s$-$s$, $s$-$p$,..., $p$-$s$, $p$-$p$,..., $d$-$d$ RSSHGTF pair $G^a_x$, $G^a_y$ and $G^a_z$ coefficients as compared to previously existing routines of the \textsc{Crystal} program.}
\label{fig:hot_stuff}
\end{figure}
\end{center}

Here the explicit expressions for the $G^{A_i}$ coefficients are predetermined using the computer algebra system (CAS) for symbolic computation available in \textsc{Matlab}, along with automated generation of \textsc{Fortran77} routines. The computational savings afforded by the new routines for $G^{A_x}$, $G^{A_y}$ and $G^{A_z}$ coefficients is documented in Fig. \ref{fig:hot_stuff}, which provides speedups of the new vs. previously existing routines for $s$ to $d$-type functions. We exclude $f$ and $g$-type functions in this presentation, as the existing routines were implemented at a later time and have different behaviours.\cite{gorbs} The speedups are asymmetric (e.g. factor of 3.66 for $d$-$p$ vs. 6.15 for $p$-$d$) because of the derivative in Eq. (\ref{eqn:RRLG}), which is only taken on the left Gaussian function. In the best cases ($p$-$d$ and $sp$-$d$), the relevant $G^{A_x}$, $G^{A_y}$ and $G^{A_z}$ coefficients are calculated over six times faster, compared to the previous implementation.

Of course, the speedups reported in Fig. \ref{fig:hot_stuff} are not reflective of the actual gains on an overall calculation, which includes more than just calculating the $G^{A_i}$ expansion coefficients of Eq. (\ref{eqn:RRLG}). In practice, an energy gradient calculation also requires a converged self-consistent field (SCF) procedure, involving i) integral calculations (in particular, evaluating the infinite Coulomb and exchange series) and their contraction with the density matrix to construct the Fock matrix in the atomic-orbital (AO) basis, followed by ii) transformation of the Fock matrix from the AO to crystalline-orbital (CO) basis, and iii) diagonalization of the CO Fock matrix. Steps i) to iii) are repeated until convergence. Once the SCF procedure is converged, the energy gradient may be subsequently computed through a procedure requiring, most importantly, the derivatives of the electron-repulsion integrals. These, in turn, are computed by a contraction of the density matrix with the $G^{A_i}$ coefficients of Eq. (\ref{eqn:RRLG}) and derivatives of the Boys' function. It is then clear that computation of the coefficients $G^{A_i}$ of Eq. (\ref{eqn:RRLG}), represents merely one (although important) step of the full calculation.

\begin{center}
\begin{figure}[b!]
\centering
\includegraphics[width=8.6cm]{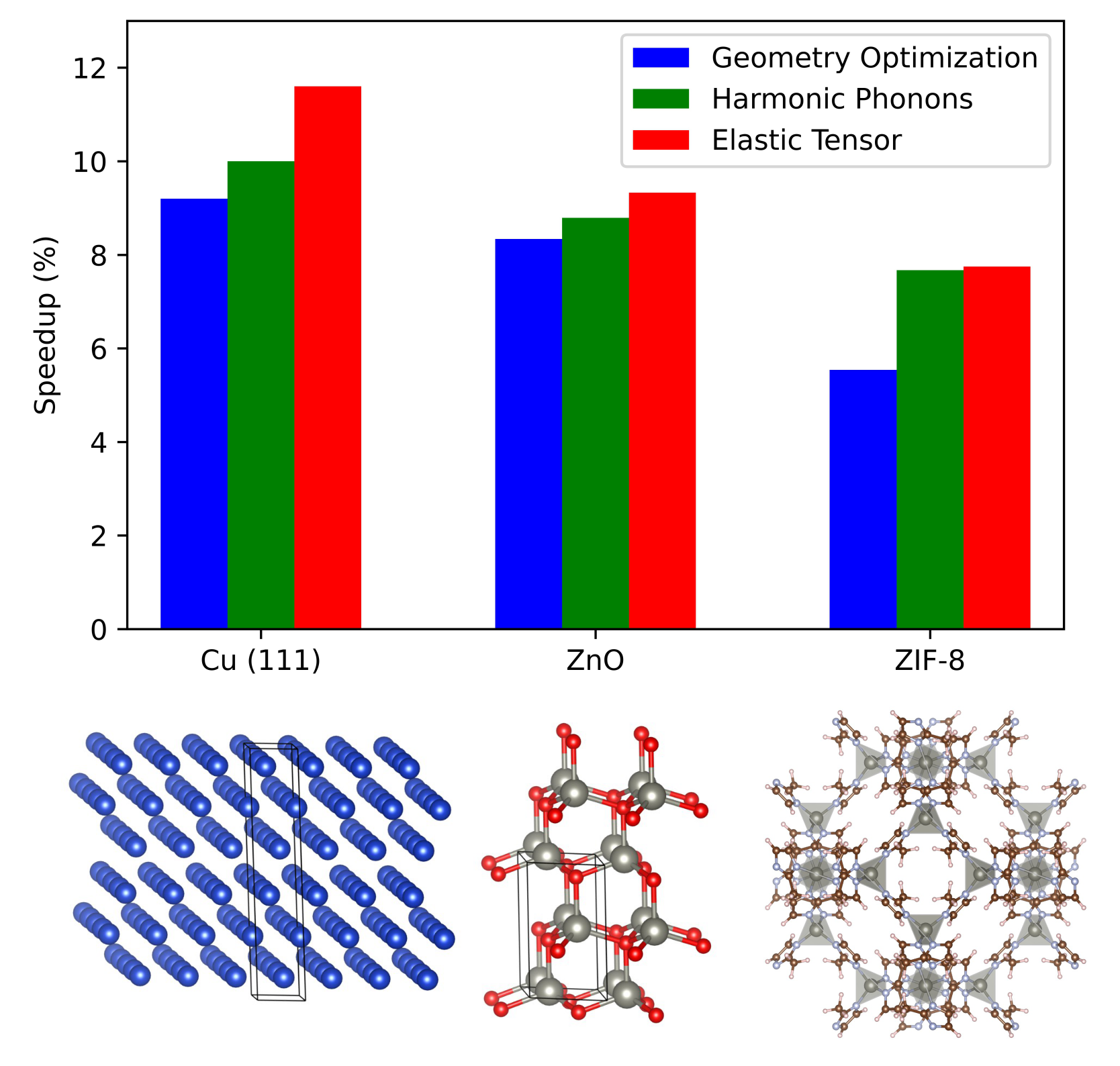}
\caption{(Upper panel) Percentage speedup of the new implementation on overall calculations (geometry optimization in blue, harmonic phonons in green, elastic tensor in red) for the three representative systems. (Lower panels) Atomic structure of the three representative systems.}
\label{fig:resu}
\end{figure}
\end{center}

In a practical calculation, the total energy gradients are used to compute a variety of physical properties of materials, including: i) the equilibrium crystal structure through a geometry optimization process, requiring first derivatives of the energy;\cite{Civalleri2001} ii) the effect of pressure on the structure via an equation-of-state or stress tensor approach, through constrained geometry optimizations;\cite{DOLL_STRESS,ELAPRES,ANISOPRES} iii) harmonic and quasi-harmonic lattice dynamics, requiring second derivatives of the energy with respect to atomic displacements, here computed as numerical first derivatives of the analytical energy gradients;\cite{freq1,freq2,PyrGro} iv) anharmonic vibrational states, requiring higher-than-quadratic terms of the potential energy surface, here computed with a finite-difference approach based on the energy and analytical first derivatives;\cite{PARTI_ANHARM,PARTII_ANHARM,ANHAPESSYM,maul2019elucidating,schireman2022anharmonic} v) Elastic and thermo-elastic constants, requiring second derivatives of the energy with respect to strain, here computed as numerical first derivatives of analytical energy gradients;\cite{ElastPerger,ElaGarnets,IntStrain,HESSvsOPT,THERMOELAST_MIO,maul2020thermoelasticity} and many others.

To provide figures that are more reflective of the actual gains of the new implementation on an actual calculation, we have performed geometry optimizations, $\Gamma$-point harmonic vibration frequency, and elastic tensor calculations on three representative systems with the \textsc{Crystal} code. All calculations are performed with all-electron basis sets and hybrid exchange-correlation functionals. The full input decks are reported in the electronic supporting information.\cite{ESI_LETT_INT}  The systems are 1) a metallic Cu (111) surface with six atoms in the primitive cell, of which three are irreducible by operations of the space group of symmetry; 2) a wide-gap semiconductor ZnO crystal with a Wurzite-type structure, with four atoms in the cell and two irreducible ones; and 3) an open-framework crystal represented by the ZIF-8 Zinc-imidazolate metal-organic framework, with 138 atoms in the cell and eight irreducible ones. Figure \ref{fig:resu} shows the atomic structure of the three systems. Symmetry is fully exploited at each step of the calculation. For each of the three systems, we repeated twice each calculation, employing the new vs. previously-existing routines for computing the RSSHGTF pair $G^a_x$, $G^a_y$ and $G^a_z$ coefficients of Eq. (\ref{eqn:RRLG}), everything else being equal.

The percentage speedup on the overall calculations is reported in the bar plot of Figure \ref{fig:resu}, being usually on the order of 10\%. In the best case (elastic tensor calculation for the dense Cu metallic surface), a speedup of about 11.6\% on the complete calculation is obtained. In the worst case (geometry optimization on the open-framework ZIF-8 crystal) a speedup of 5.54\% is reported. The gains are largest where calculation of integrals dominates over diagonalization and AO-to-CO transformation of the Fock matrix, and where calculation of energy gradients dominates over the cost of the SCF procedure. This is expected to occur in relatively small (in terms of irreducible atoms in the cell) and dense periodic systems with small or vanishing gaps (in this case, represented by the Cu metallic surface). Inspection of the figure suggests that the speedup systematically increases when moving from a geometry optimization to harmonic phonon or elastic tensor calculations. The latter differ from the former in one significant respect: they involve many calculations at low symmetry nuclear configurations (either atomically displaced or strained), which suggests that the relative cost associated to the calculation of the forces increases upon symmetry removal and thus makes the new implementation particularly advantageous for low symmetry systems.

\subsection{Second-Order Derivatives with respect to Atomic Positions}

One particular nice feature of the present symbolic approach is its straightforward generalization to other derivatives. We sketch this first for the computation of second derivatives of the integrals w.r.t. nuclear displacements. Taking the derivative of Eq. (\ref{eqn:RRLG}) with respect to a pair of arbitrary centers $I_i, J_j = A_x,A_y,A_z,B_x,B_y,B_z$, we obtain:
\begin{widetext}
\begin{eqnarray}
\label{eqn:RRLGG}
\frac{\partial}{\partial I_i} \frac{\partial}{\partial J_j} R \left( \alpha,\mathbf{r}-\mathbf{A},n,l,m_l\right) R \left( \beta,\mathbf{r}-\mathbf{B},n^\prime,l^\prime,m_l^\prime\right) &=&
\frac{\partial}{\partial I_i}  \sum_{t,u,v}^{\mathcal{G} \left( n,l,n^\prime,l^\prime \right) } G^{J_j}_{t,u,v} \left[ n,l,m_l,n^\prime,l^\prime,m_l^\prime\right] \Lambda_{t,u,v} \left( \gamma,\mathbf{r}-\mathbf{P}\right) \nonumber \\
&\equiv& \sum_{t,u,v}^{\mathcal{F} \left( n,l,n^\prime,l^\prime \right)  } F^{I_i J_j}_{t,u,v} \left[ n,l,m_l,n^\prime,l^\prime,m_l^\prime\right] \Lambda_{t,u,v} \left( \gamma,\mathbf{r}-\mathbf{P}\right) \; .
\end{eqnarray}
\end{widetext}
It will become apparent below that the set $\mathcal{F} \left( n,l,n^\prime,l^\prime \right)$ includes all positive integer triplets that satisfy $t+u+v \le 2n +2n^\prime + l + l^\prime+2$.  Distributing the derivative in Eq. (\ref{eqn:RRLGG}), gives:
\begin{widetext}
\begin{eqnarray}
\label{eqn:RRLGG_dist}
\sum_{t,u,v}^{\mathcal{F} \left( n,l,n^\prime,l^\prime \right)  } F^{I_i J_j}_{t,u,v} \left[ n,l,m_l,n^\prime,l^\prime,m_l^\prime\right] \Lambda_{t,u,v} \left( \gamma,\mathbf{r}-\mathbf{P}\right) &=& 
\sum_{t,u,v}^{\mathcal{G} \left( n,l,n^\prime,l^\prime \right)} \frac{\partial}{\partial I_i}  G^{J_j}_{t,u,v} \left[ n,l,m_l,n^\prime,l^\prime,m_l^\prime\right] \Lambda_{t,u,v} \left( \gamma,\mathbf{r}-\mathbf{P}\right) \nonumber \\
&+& \frac{\zeta_I}{\gamma} G^{J_j}_{t,u,v} \left[ n,l,m_l,n^\prime,l^\prime,m_l^\prime\right] \Lambda_{t+\delta_{i,x},u+\delta_{i,y},v+\delta_{i,z}} \left( \gamma,\mathbf{r}-\mathbf{P}\right) \; ,
\end{eqnarray}
\end{widetext}
where $\zeta_I=\alpha$ if $I=A$ and $\zeta_I=\beta$ if $I=B$. From Eq. (\ref{eqn:RRLGG_dist}), we deduce:
\begin{eqnarray}
\label{eqn:RRLGG_done}
F^{I_i J_j}_{t,u,v} \left[ n,l,m_l,n^\prime,l^\prime,m_l^\prime\right] = \frac{\partial}{\partial I_i}  G^{J_j}_{t,u,v} \left[ n,l,m_l,n^\prime,l^\prime,m_l^\prime\right] \nonumber \\
+ \frac{\zeta_I}{\gamma} G^{J_j}_{t-\delta_{i,x},u-\delta_{i,y},v-\delta_{i,z}} \left[ n,l,m_l,n^\prime,l^\prime,m_l^\prime\right] \; .
\end{eqnarray}
From Eq. (\ref{eqn:RRLGG_done}), we obtain the important result that once the symbolic expressions for the $G^{J_j}$ are known, the
ones for the second energy gradients $F^{I_i J_j}$ can be trivially obtained from symbolic differentiation and addition.

\subsection{First-Order Derivatives with respect to a Magnetic Field}

Another noteworthy and straightforward generalization of the present approach is the computation of first energy derivatives w.r.t. an applied magnetic field $\boldsymbol{\mathcal{B}}$. Then, with a finite basis-set, the well-known gauge-origin problem is typically solved by including field-dependent phase factors in the basis functions - the so-called gauge-including atomic-orbital, or GIAO, approach:\cite{cheeseman2000hartree,stephens2001calculation,ruud2002gauge}
\begin{equation}
\label{eqn:GIAO}
\tilde{R} \left( \alpha,\mathbf{r}-\mathbf{A},n,l,m_l\right) = e^{-\frac{\i}{2} \boldsymbol{\mathcal{B}} \wedge \mathbf{A} \cdot \mathbf{r} }  R \left( \alpha,\mathbf{r}-\mathbf{A},n,l,m_l\right) \; .
\end{equation}
For the purposes of computing integrals for magnetic response properties that are first order in the field, the RSSHGTF pair-product is corrispondingly modified as:\cite{krykunov2006calculation}
\begin{equation}
\frac{\i}{2} \mathbf{r}  \wedge  \left( \mathbf{B} - \mathbf{A}  \right)  R \left( \alpha,\mathbf{r}-\mathbf{A},n,l,m_l\right) R \left( \beta,\mathbf{r}-\mathbf{B},n^\prime,l^\prime,m_l^\prime \right)  \nonumber \; .
\end{equation}
Then, considering terms, for instance, involving $r_x$, the RSSHGTF pair-product may be expanded as (up to a constant factor):
\begin{eqnarray}
\label{eqn:GIAO_E}
r_x R \left( \alpha,\mathbf{r}-\mathbf{A},n,l,m_l\right) R \left( \beta,\mathbf{r}-\mathbf{B},n^\prime,l^\prime,m_l^\prime \right) \nonumber \\
\equiv \sum_{t,u,v}^{\tilde{\mathcal{E}} \left( n,l,n^\prime,l^\prime \right)} \tilde{E}_{t,u,v}^{(x)} \left[ n,l,m_l,n^\prime,l^\prime,m_l^\prime \right] \Lambda_{t,u,v} \left( \gamma,\mathbf{r}-\mathbf{P}\right) \; .
\end{eqnarray}
As will become apparent below, the set $\tilde{\mathcal{E}} \left( n,l,n^\prime,l^\prime \right)$ coincides with $\mathcal{G} \left( n,l,n^\prime,l^\prime \right)$. We now make use of the following recurrence relation for HGTF:\cite{Saund,desmarais2020development}
\begin{eqnarray}
\label{eqn:hgtf_rec}
r_x \Lambda_{t,u,v} \left( \gamma,\mathbf{r}-\mathbf{P}\right) = \frac{1}{2 \gamma} \Lambda_{t+1,u,v} \left( \gamma,\mathbf{r}-\mathbf{P}\right) \nonumber \\
+ P_x \Lambda_{t,u,v} \left( \gamma,\mathbf{r}-\mathbf{P}\right) + t \Lambda_{t-1,u,v} \left( \gamma,\mathbf{r}-\mathbf{P}\right) \; .
\end{eqnarray}
Then, inserting Eq. (\ref{eqn:hgtf_rec}) into Eq. (\ref{eqn:GIAO_E}), and using also Eq. (\ref{eqn:RRL}), gives:
\begin{widetext}
\begin{eqnarray}
\sum_{t,u,v}^{\tilde{\mathcal{E}} \left( n,l,n^\prime,l^\prime \right)} \tilde{E}_{t,u,v}^{(x)} \left[ n,l,m_l,n^\prime,l^\prime,m_l^\prime \right] \Lambda_{t,u,v} \left( \gamma,\mathbf{r}-\mathbf{P}\right) 
&=& \sum_{t,u,v}^{\mathcal{E} \left( n,l,n^\prime,l^\prime \right)}  E_{t,u,v} \left[ n,l,m_l,n^\prime,l^\prime,m_l^\prime \right] \Big\{ \frac{1}{2 \gamma} \Lambda_{t+1,u,v} \left( \gamma,\mathbf{r}-\mathbf{P}\right) \nonumber \\
&+& P_x \Lambda_{t,u,v} \left( \gamma,\mathbf{r}-\mathbf{P}\right) + t \Lambda_{t-1,u,v} \left( \gamma,\mathbf{r}-\mathbf{P}\right) \Big\} \; ,
\end{eqnarray}
\end{widetext}
from which we deduce the relation:
\begin{widetext}
\begin{eqnarray}
\tilde{E}_{t,u,v}^{(x)} \left[ n,l,m_l,n^\prime,l^\prime,m_l^\prime \right] = \frac{1}{2 \gamma} E_{t-1,u,v} \left[ n,l,m_l,n^\prime,l^\prime,m_l^\prime \right] 
+ P_x E_{t,u,v} \left[ n,l,m_l,n^\prime,l^\prime,m_l^\prime \right] + (t+1) E_{t+1,u,v} \left[ n,l,m_l,n^\prime,l^\prime,m_l^\prime \right] \nonumber
\end{eqnarray}
\end{widetext}
and therefore once the symbolic expressions are known for $E$, the first-order GIAO coefficients $\tilde{E}_{t,u,v}^{(x)}$ may also be obtained by elementary symbolic manipulation. The procedure can also be extended to derivatives of higher order in the field, using the methods provided above.

\section{Conclusions}
A computational procedure was developed for the efficient calculation of derivatives of integrals over non-separable Gaussian-type basis functions, within the framework of Saunders' algorithm. The strategy involved symbolic computation with computer algebra systems, as well as automated generation of optimized subroutines and took full advantage of sparsity. The procedure was practically applied to calculating first energy derivatives with respect to nuclear displacements and lattice parameters of molecules and materials. The implementation in the \textsc{Crystal} code considerably improved computational efficiency over the previous one. The facility in generalizing the proposed symbolic approach to other derivatives was noted, and two generalizations of particular future interest were illustrated.

\section*{Acknowledgements}

J.K.D. is grateful to the National Science and Engineering Research Council of the Government of Canada for a Postdoctoral fellowship application No. 545643.

\end{document}